\documentclass{iaus}
\usepackage{graphicx}

\newcommand{\Oiii}{[{\sc O$\,$iii}]}

\title[PNe with SAURON]{The central PNe populations of external
  galaxies with SAURON}

\author[Marc Sarzi]{Marc Sarzi}

\affiliation{Centre for Astrophysics Research, University of Hertfordshire,
College Lane, Hatfield, Herts, AL10 9AB, UK\\email: {\tt m.sarzi@herts.ac.uk}}

\pubyear{2008}
\volume{xxx}  
\pagerange{119--126}
\jname{Title of your IAU Symposium}
\editors{A.C. Editor, B.D. Editor \& C.E. Editor, eds.}
\begin{document}

\maketitle

\begin{abstract}

Thanks to {\tt SAURON\/} integral-field observations we uncovered the
Planetary Nebulae (PNe) populations inhabiting the central and nuclear
regions of our galactic neighbours M32 and M31, respectively, and
discuss the significant differences between their corresponding PNe
luminosity functions in light of the properties of their parent
stellar populations. In particular, we conclude that the lack of
bright PNe in the nuclear regions of M31 is likely linked to the
nearly Solar value for the stellar metallicity, consistent with
previous suggestions that a larger metallicity would bias the
Horizontal-Branch (HB) populations toward bluer colors, with fewer red
HB stars capable of producing PNe and more blue HB stars that instead
could contribute to the far-UV flux that is observed in metal-rich
early-type galaxies and, incidentally, also in the nucleus of M31.
\end{abstract}

\firstsection 
\section{Introduction}

Planetary Nebulae (PNe) in external galaxies are mostly regarded
either as tracers of the gravitational potential (e.g., Romanowsky et
al. 2003) or as indicators for the distance of their galactic hosts
(e.g., Ciardullo et al. 1989), with the latter advantage owing to the
nearly universal -- though not fully understood -- shape of the PNe
luminosity function (PNLF, generally in the \Oiii$\lambda5007$
emission).
Yet extra-galactic PNe can also be used as probes of their parent
stellar population (see, e.g., Ciardullo 2006) and understanding in
particular the origin of the PNLF is a puzzle that, once solved,
promises to reveal new clues on the late stages of stellar evolution
and on the formation of PNe themselves (e.g., Ciardullo et al. 2005;
Buzzoni, Arnaboldi \& Corradi 2006).

PNe originates from horizontal-branch (HB) stars that climb back the
asymptotic giant branch (AGB) at the end of their helium-burning
phase, when these stars leave the AGB and quickly cross the
Hertzprung-Russell diagram on their way towards the cooling track of
white dwarves (WD).
For a population with a given age and metallicity, HB stars have
nearly the same helium core mass ($\sim 0.5\,M_{\odot}$) but a range
of hydrogen shell mass ($\sim 0.001 - 0.3 \,M_{\odot}$), with the
reddest stars having also the largest H-shells and originating from
the most massive main-sequence progenitors.
Only HB stars with a considerable H-shell ascend toward the AGB and
eventually lead to the formation of a PN, whereas the bluest HB stars
with little envelope mass head straight toward the WD cooling curve by
evolving first to higher luminosities and effective temperatures (the
so-called AGB-manqu\'e phase).

According to this simple picture, galaxies with on-going star
formation should show brighter PNe than quiescent systems where
massive stars have long disappeared (e.g., Marigo et al. 2004), but in
fact the PNLF of young and old galaxies are relatively similar.
In particular, all extra-galactic PNe surveys indicate a common and
bright cut-off for the PNLF, which led Ciardullo et al. (2005) to
suggest a binary evolution for the progenitors of the brightest PNe
that would be common to different kind of galaxies.
If galaxies seem to invariably host very bright PNe, their specific
content of PNe - that is the number of PNe normalised by a galaxy
bolometric luminosity - appears to vary with the metallicity of the
stellar population. More specifically, Buzzoni, Arnaboldi \& Corradi
(2006) found that more metal rich galaxies show comparably fewer PNe,
which also corresponds to larger far-UV fluxes.
Interestingly, this may indicate that at a given mean stellar age, a
larger metallicity would bias the HB population towards fewer stars
with massive H-shell capable to lead to the formation of PNe, with a
larger fraction of blue HB stars that contribute instead to the
overall far-UV flux of their host galaxy (i.e. the so-called
UV-upturn, Burnstein et al. 1988) as they follow their AGB-manqu\'e
evolution.

Within this context, we note that whereas our knowledge of both the
shape and normalisation of the PNLF comes chiefly from the peripheral
PNe populations of galaxies, both measurements for the stellar
metallity and the UV spectral shape of galaxies pertain to their
optical regions.
This is because narrow-band imaging or slitless spectroscopy - the
most common techniques employed to find extragalactic PNe - find it
hard to detect PNe against a strong stellar background, whereas
measuring the strength of stellar absorption lines or imaging the far-UV
flux of galactic halos is prohibitively expensive in terms of
telescope time.
Such a dramatic spatial inconsistency needs to be resolved if we ought
to understand the link between PNe and the properties of their parent
stellar populations, in particular if we consider that such a
connection may already not be entirely within our grasp, as Hubble
Space Telescope (HST) observations for the UV color-magnitude diagram
of M32 suggests (Brown et al. 2008).

\section{PNe with Integral-Field Spectroscopy}

Integral-field spectroscopy (IFS) can overcome the previous technical
limitations and shed more light on the link between PNe and their
parent stellar population by allowing to both explore the central PNe
population of galaxies and measure the stellar metallicity of their
halos.
Indeed, IFS makes it possible to carefully model the central
stellar spectrum of a galaxy and thus reveal the presence of PNe
deeply embedded in it. At the same time, adding up all the spectra
obtained by an integral-field unit effectively turns it into a large
light bucket that allows to measure the strength of stellar absorption
lines out to large galactic radii.
For instance, using the {\tt SAURON\/} integral-field spectrograph
mounted on the 4m William Herschel telescope, Sarzi et al. (2011) more
than doubled the number of PNe known in the optical regions of M32 in
just 20 minutes of observations, whereas Weijmans et al. (2009) could
map the stellar age and metallicity of NGC~821 and NGC~3379 out to 3
and 4 effective radii, respectively, by spending only few hours per
galaxy.

\begin{figure}[b]
  \begin{center}
    \includegraphics[width=0.95\textwidth]{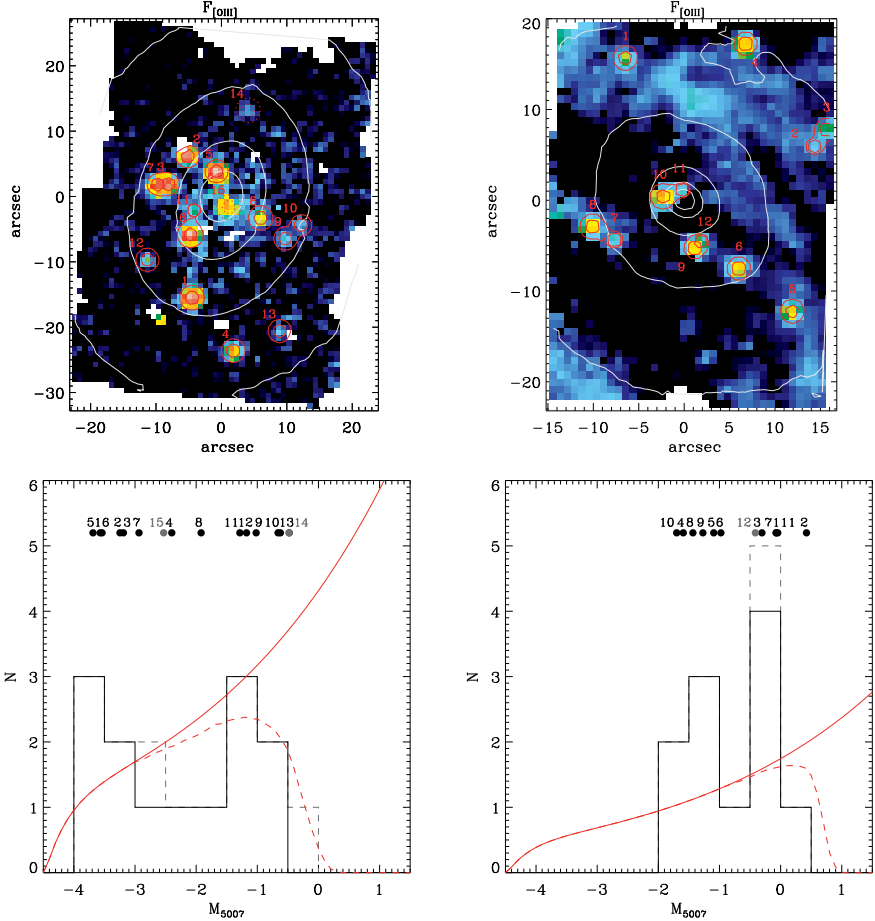} 
    \caption{{\tt SAURON\/} maps for the flux of the \Oiii$\lambda
      5007$ emission (upper panels) from the central regions of M32
      (left) and the nuclear regions of M31 (right) together with the
      luminosity function of the PNe found in them (lower panels). In
      the \Oiii\ maps the white contours outline the galaxy stellar
      isophotes whereas the position of the detected PNe is shown by
      the red contours.
      The absolute magnitude of these PNe can be read in the lower
      panels, where the red solid and dashed lines also show the
      theoretical form of the PNLF (from Ciardullo et al. 1989) as it
      is and after accounting for the incompletness of our
      observations, respectively.
      Grey points and dashed histograms correspond to marginal PNe
      detections.
      In the case of M32, the models have been normalised to match the
      observed number of PNe to the number of objects that we would
      expect to detect according to the incompleteness-corrected PNLF
      model. For M31 the nuclear PNe population is unlikely to have
      been drawn from such a PNLF model, and the lines are shown
      mostly for an illustrative purpose and are normalised by simply
      assuming the same specific PNe density of M32.}
    \label{fig1}
  \end{center}
\end{figure}

\section{Comparing the Central PNe Populations of M32 and M31}

Following the work of Sarzi et al. for the central regions of M32,
Pastorello et al. (2012, in preparation) carried a similar analysis of
the nuclear regions of M31, further showing how IFS can detect PNe
also in the presence of diffuse ionised-gas emission (as is often
observed in early-type galaxies) and that for this purpose the {\tt
  SAURON\/} data are just as good as narrow-band HST images of much
higher spatial resolution.

Fig.~1 shows both the PNe detected by Pastorello et al. in the central
30pc of M31 and those found by Sarzi et al. in the optical regions of
M32 (within one effective radius $R_e$), while also comparing their
corresponding luminosity functions to the shape that the theoretical
PNLF of Ciardullo et al. (1989) takes after accounting for the
incompleteness of our observations (red dashed lines).
Already at first glance the PNe populations of M32 and M31 appear to
be different, with the PNLF of M31 looking hardly consistent with the
theoretical expectations and deficient in bright PNe.
In fact, whereas a Kolgomorov-Smirnov test indicates that the
incompleteness-corrected theoretical PNLF of Ciardullo et al. can be
regarded as the parent distribution for the PNe found in the optical
regions of M32 at an 82\% confidence level, the same test returns only
a 20\% chance of that happening for the nuclear PNe of M31.
Furthermore, even assuming that the PNe of M31 were drawn from such a
standard PNLF, simulations like those presented by Sarzi et al.  would
generate synthetic PNLFs with no bright PNe at all (i.e., within 2.5
magnitudes of the bright cutoff at $M_{5007}=-4.47$) only in 8\% of
the cases.
It is therefore unlikely that the lack of bright PNe in the nuclear
regions of M31 (confirmed also by the HST observations presented at
this conference by Girardi et al.) is merely due to bad luck.

\section{Connection with the Parent Stellar Population}

It is interesting to consider such remarkable differences between the
central and nuclear PNLF of M32 and M31, respectively, in light of the
properties of the parent stellar population of such PNe systems.

Our {\tt SAURON\/} observations encompass nearly the same amount of
stellar light in these two galaxies and in both cases the stellar
population is fairly old.
However, the nuclear population of M31 is more metal rich than that of
the central regions of M32.
More specifically, our own {\tt SAURON\/} stellar absorption-line
measurements indicate a nearly Solar value of $\rm [Fe/H] \sim -0.2$
for the average metallicity of the nuclear stars of M31, in line with
what found in the centers of low mass early-type galaxies (e.g., as in
{\tt SAURON\/} survey; Kuntschner et al. 2006), whereas the optical
regions of M32 display stellar metallicity values around $\rm [Fe/H]
\sim -0.5$, well below Solar standards and closer to what observed in
the outskirts of more typical early-type galaxies (see, e.g., Weijmans
et al. 2009).
At the same time, is has long been known since the first
low-resolution UV spectroscopic observations of Burstein et al. (1988)
with the International Ultraviolet Explorer (IUE) that M31 shows
stronger far-UV fluxes than M32 (within their central $\sim 20''$).

That the lack of bright PNe in the nuclear regions of M31 coincides
with the presence of a UV-upturn and with a larger stellar metallicity
than in the case of M32 supports the scenario advanced by Buzzoni,
Arnaboldi \& Corradi, whereby at a given mean age a more metal-rich
stellar population would display a bluer HB population with fewer red
HB stars capable of producing PNe and a larger fraction of blue HB
stars that during their AGB-manqu\'{e} phase would contribute to
larger far-UV fluxes.
Yet, our data further indicate, as intuition suggests, that such a HB
bias would start by suppressing the reddest HB stars that lead to the
brighest PNe, thus leading also to a change in the shape of the PNLF.
It would be interesting to see if even more metal-rich stellar systems
show such a faint PNLF cutoff, as found for the first time by our {\tt
  SAURON\/} observations in the nuclear region of M31 (see Pastorello
et al., for details).

\end{document}